\documentclass{jpp}
\usepackage{graphicx,natbib,bm,url,color}
\graphicspath{{./fig/}{./png/}}

\newcommand{\EQ}{\begin{equation}}
\newcommand{\EN}{\end{equation}}
\newcommand{\EQA}{\begin{eqnarray}}
\newcommand{\ENA}{\end{eqnarray}}

\newcommand{\Eq}[1]{equation~(\ref{#1})}

\newcommand{\App}[1]{Appendix~\ref{#1}}

\newcommand{\Sec}[1]{section~\ref{#1}}

\newcommand{\Fig}[1]{figure~\ref{#1}}
\newcommand{\FFig}[1]{Figure~\ref{#1}}

\newcommand{\Tab}[1]{table~\ref{#1}}

\newcommand{\bra}[1]{\langle #1\rangle}

{}
{}

{}
{}

{}
{}
{}
{}
{}
{}
{}
{}
{}
{}
{}
{}
{}
{}

{}

{}

{}
{}
{}

\newcommand{\zzz}{\hat{\mbox{\boldmath $z$}} {}}

\newcommand{\xx}{\bm{x}}

\newcommand{\rr}{\bm{r}}

\newcommand{\BB}{\bm{B}}

\newcommand{\JJ}{\bm{J}}

\newcommand{\AAA}{\bm{A}}

\newcommand{\uu}{\bm{u}}

\newcommand{\nab}{{\bm{\nabla}}}

\newcommand{\SSSS}{\mbox{\boldmath ${\sf S}$} {}}

\newcommand{\DD}{{\rm D} {}}

\newcommand{\dd}{{\rm d} {}}

\def\degr{\hbox{$^\circ$}}

\def\Sp{\mbox{\rm Sp}}

\def\Pm{\mbox{\rm Pr}_{\rm M}}

\def\Lu{\mbox{\rm Lu}}

\def\tauA{\tau_{\rm A}}

\def\EEM{{\cal E}_{\rm M}}

\def\EM{E_{\rm M}}

\def\cs{c_{\rm s}}

\def\CM{C_{\rm M}}

\def\xiM{\xi_{\rm M}}
\def\xiT{\xi_{\rm T}}

\def\vA{v_{\rm A}}

\def\vAz{v_{\rm A0}}

\def\kf{k_{\rm f}}
\def\tp{t_{\rm p}}

\def\EM{E_{\rm M}}

\def\Brms{B_{\rm rms}}

\def\Jrms{J_{\rm rms}}

\title{Inverse cascade from helical and nonhelical decaying columnar magnetic fields}

\author[A. Brandenburg et al.]{
Axel Brandenburg\aff{1,2,3,4}\corresp{\email{brandenb@nordita.org}},
Longqing Yi\aff{5,6}\corresp{\email{lqyi@sjtu.edu.cn}}
\, \and \,
Xianshu Wu\aff{5}
}
\affiliation{
\aff{1}Nordita, KTH Royal Institute of Technology and Stockholm University,
Hannes Alfv\'ens v\"ag 12, SE-10691 Stockholm, Sweden
\aff{2}The Oskar Klein Centre, Department of Astronomy,
Stockholm University, AlbaNova, SE-10691 Stockholm, Sweden
\aff{3}McWilliams Center for Cosmology \& Department of Physics,
Carnegie Mellon University, Pittsburgh, PA 15213, USA
\aff{4}School of Natural Sciences and Medicine, Ilia State University,
3-5 Cholokashvili Avenue, 0194 Tbilisi, Georgia
\aff{5}Tsung-Dao Lee Institute, Shanghai Jiao Tong University, Shanghai 201210, China
\aff{6}School of Physics and Astronomy, Shanghai Jiao Tong University, Shanghai 200240, China
}

\date{\today}

\begin{document}

\maketitle

\begin{abstract}
Powerful lasers may in future produce magnetic fields that would allow us to study turbulent magnetohydrodynamic inverse cascade behavior.
This has so far only been seen in numerical simulations.
In the laboratory, however, the produced fields may be highly anisotropic.
Here, we present corresponding simulations to show that, during the turbulent decay, such a magnetic field undergoes spontaneous isotropization.
As a consequence, we find the decay dynamics to be similar to that in isotropic turbulence.
We also find that an initially pointwise nonhelical magnetic field is unstable and develops magnetic helicity fluctuations that can be quantified by the Hosking integral.
It is a conserved quantity that characterizes magnetic helicity fluctuations and governs the turbulent decay when the mean magnetic helicity vanishes.
As in earlier work, the ratio of the magnetic decay time to the Alfv\'en time is found to be around $50$ in the helical and nonhelical cases.
At intermediate times, the ratio can even reach a hundred.
This ratio determines the endpoints of cosmological magnetic field evolution.
\end{abstract}

\section{Introduction}

In the absence of any initial velocity field and without any type of forcing, an initially random magnetic field can only decay.
This decay can be sped up by turbulent gas motions driven through the Lorentz force that is being exerted by the magnetic field itself.
The decay of such a random field obeys powerlaw behavior where the magnetic energy density $\EEM$ decays with time $t$ like $\EEM(t)\propto t^{-p}$,
and the magnetic correlation length $\xiM$ increases like $\xiM\propto t^q$.
For a helical magnetic field, we have $p=q=2/3$ \citep{Hat84,BM99}, while for a nonhelical magnetic field, we have $p=10/9$ and $q=4/9$ \citep{HS21,Zhou+22}.
Such a decay has been seen in many hydromagnetic numerical simulations \citep{BKT15,HS21,Armua+23,BSV23}, but not yet in plasma experiments.
With the advance of high-powered lasers it is already possible to produce magnetic fields in the laboratory \citep{Tzeferacos+18},
and similar advances may also allow us to achieve sufficient scale separation to perform meaningful inverse cascade experiments.
However, such magnetic fields may be strongly anisotropic, so the question arises to what extent this affects the otherwise familiar decay dynamics.

Our goal here is to study the decay of an array of magnetic flux tubes with an electric current that is aligned with the magnetic field \citep{Jiang+21}.
Such a field is indeed highly anisotropic such that the correlation length in the direction along the tubes is much larger than that perpendicular to it.
A simple numerical realization of such a magnetic field is what is called the Roberts field I, which is more commonly also known as Roberts flow I.
It is one of four flow fields studied by \cite{Rob72} in the context of dynamo theory.
The field is fully helical, but with a slight modification, it can become a pointwise nonhelical field, which is then called the Roberts field II.
Both fields are here of interest.
They are defined in \Sec{OurModel}, along with a proper measure of anisotropy, the relevant evolution equations, and relevant input and output parameters.
In \Sec{Results}, we present numerical results for both flows using different magnetic diffusivities and scale separation ratios.
Inverse cascading during the turbulent decay of helical and nonhelical magnetic fields has applications to primordial magnetic fields in the radiation dominated era of the Universe,
which are discussed in \Sec{CosmologicalApplications}.
We conclude in \Sec{Conclusions}.

\section{Our model}
\label{OurModel}

\subsection{Roberts fields}

To fix our geometry, we assume magnetic flux tubes to extend in the $z$ direction and being perpendicular to the $xy$ plane.
Such a field can be realized by the so-called Roberts field I, i.e., the magnetic field $\BB$ is given by
\begin{equation}
\BB=\BB_\mathrm{I}\equiv\nab\times\phi\zzz+\sqrt{2}k_0\phi\zzz, \quad \mbox{where}\quad \phi=k_0^{-1} B_0 \sin k_0 x\sin k_0 y.
\end{equation}
is an $xy$ periodic field.
Such a magnetic field has a component in the $z$ direction, but no variation along that direction, so it is highly anisotropic.
This may change with time as the magnetic field undergoes a turbulent decay.
The Roberts field I is maximally helical with $\AAA\cdot\BB=\sqrt{2} k_0^{-1} B_0^2 (\sin^2 k_0x+\sin^2 k_0y)$, so $\bra{\AAA\cdot\BB}=\sqrt{2} k_0^{-1} B_0^2$.
Here, $\AAA$ is the magnetic vector potential and $\BB=\nab\times\AAA$.
The Roberts field II, by contrast, is given by
\begin{equation}
\BB=\BB_\mathrm{II}\equiv\nab\times\phi\zzz+\kf\tilde{\phi}\zzz, \quad \mbox{where}\quad \tilde{\phi}=k_0^{-1} B_0 \cos k_0 x\cos k_0 y,
\end{equation}
where $\tilde{\phi}$ is $90\degr$ phase shifted in the $x$ and $y$ directions relative to $\phi(x,y)$ and
$\kf=\sqrt{2}k_0$ is the eigenvalue of the curl operator for field~I, i.e., $\nab\times\BB_\mathrm{I}=\kf\BB_\mathrm{I}$,
so $\BB_\mathrm{I}\cdot\nab\times\BB_\mathrm{I}=\kf\BB_\mathrm{I}^2$, while $\BB_\mathrm{II}\cdot\nab\times\BB_\mathrm{II}=0$ pointwise.
Both for fields~I and II, we have $\bra{\BB^2}=2B_0^2$.

\subsection{Quantifying the emerging anisotropy}

To quantify the degree of anisotropy, we must separate the derivatives of the magnetic field along the $z$ direction ($\nab_\|$)
from those perpendicular to it ($\nab_\perp$), so $\nab=\nab_\|+\nab_\perp$.
We also decompose the magnetic field analogously, i.e., $\BB=\BB_\|+\BB_\perp$.
The mean current density can be decomposed similarly, i,e., $\JJ=\JJ_\|+\JJ_\perp$, but this decomposition mixes the underlying derivatives.
We see this by computing $\JJ\equiv\nab\times\BB$ (where the permeability has been set to unity).
Using this decomposition, we find
\begin{equation}
\JJ=\nab_\|\times\BB_\perp+\nab_\perp\times\BB_\|+\nab_\perp\times\BB_\perp,
\label{Jdecomp}
\end{equation}
noting that $\nab_\|\times\BB_\|=0$.
The term of interest for characterizing the emergent isotropization is the first one, $\nab_\|\times\BB_\perp$, because it involves only parallel derivatives ($z$ derivatives), which vanish initially.
We monitor the ratio of its mean squared value to $\bra{\JJ^2}$.

The last term in \Eq{Jdecomp} is just $\JJ_\|=\nab_\perp\times\BB_\perp$, but the first and second terms cannot simply be expressed in terms of $\JJ_\perp$,
although $\nab_\|\times\BB_\perp$ would be $\JJ_\perp$ if the magnetic field only had a component in the plane,
and $\nab_\perp\times\BB_\|$ would be $\JJ_\perp$ if the magnetic field only had a component out of the plane.
We therefore denote those two contributions in the following by $\JJ_{\perp\perp}$ and $\JJ_{\perp\|}$, respectively, so that $\JJ_{\perp\perp}+\JJ_{\perp\|}=\JJ_\perp$.

Thus, as motivated above, to monitor the emergent isotropization, we determine $\bra{\JJ_{\perp\perp}^2}/\bra{\JJ^2}$.
For isotropic turbulence, we find that this ratio is about $4/15\approx0.27$, and this is also true for $\bra{\JJ_{\perp\|}^2}/\bra{\JJ^2}$; see \App{IsotropicTurbulence} for an empirical demonstration.
In the expression for $\bra{\JJ^2}$, there is also a mixed term, $\JJ_{\perp\textrm{m}}^2=-2\bra{B_{x,z}B_{z,x}+B_{y,z}B_{z,y}}$, which turns out to be positive in practice.
Here, commas denote partial differentiation.
Thus, we have
\begin{equation}
\bra{\JJ^2}=
\bra{\JJ_{\perp\perp}^2}+
\bra{\JJ_{\perp\|}^2}+
\bra{\JJ_{\perp\textrm{m}}^2}+
\bra{\JJ_{\|}^2}.
\label{Jall}
\end{equation}
In the isotropic case, we find $\bra{\JJ_{\|}^2}/\bra{\JJ^2}=1/3$, and for the mixed term we then have $\bra{\JJ_{\perp\textrm{m}}^2}/\bra{\JJ^2}=2/15\approx0.13$.

\subsection{Evolution equations}

To study the decay of the magnetic field, we solve the evolution equations of magnetohydrodynamics (MHD) for an isotropic compressible gas with constant sound speed $\cs$,
so the gas density $\rho$ is proportional to the pressure $p=\rho\cs^2$.
In that case, $\ln\rho$ and the velocity $\uu$ obey
\begin{equation}
\frac{\DD\ln\rho}{\DD t}=-\nab\cdot\uu,
\label{DlnrhoDt}
\end{equation}
\begin{equation}
\frac{\DD\uu}{\DD t}=-\cs^2\nab\ln\rho+\frac{1}{\rho}\left[\JJ\times\BB+\nab\cdot(2\rho\nu\SSSS)\right],
\label{DuDt}
\end{equation}
where $\DD/\DD t=\partial/\partial t+\uu\cdot\nab$ is the advective derivative, $\nu$ is the kinematic viscosity, $\SSSS$ is the rate-of-strain tensor
with components $\mathsf{S}_{ij}=(u_{i,j}+u_{j,i})/2-\delta_{ij}\nab\cdot\uu/3$, and $\alpha$ is the photon drag coefficient \citep{BJ04}, which is included in some of our simulations.
To ensure that the condition $\nab\cdot\BB=0$ is obeyed at all times, we also solve the uncurled induction equation for $\AAA$, i.e.,
\begin{equation}
\frac{\partial\AAA}{\partial t}=\uu\times\BB-\eta\JJ.
\label{dAdt}
\end{equation}
As before, the permeability is set to unity, so $\JJ=\nab\times\BB$ is the current density.


We use the \textsc{Pencil Code} \citep{JOSS}, which is well suited for our MHD simulations.
It uses sixth order accurate spatial discretizations and a third order time-stepping scheme.
We adopt periodic boundary conditions in all three directions, so the mass is conserved and the mean density is $\bra{\rho}\equiv\rho_0$ is constant.
The size of the domain is $L_\perp\times L_\perp\times L_\|$ and the lowest wavenumber in the plane is $k_1=2\pi/L_\perp$.
By default, we choose $\rho_0=k_1=\cs=\mu_0=1$ which fixes all dimensions in the code.

\subsection{Input and output parameters}

In the following, we study cases with different values of $k_0$.
We specify the amplitude of the vector potential to be $A_0=0.02$ for most of the runs with Roberts field I and $A_0=0.05$ for Roberts field II.
We use $k_0=16$, so $B_0=k_0A_0=0.32$ for field~I and $0.8$ for field~II.
For other values of $k_0$, we adjust $A_0$ such that $B_0$ is unchanged in all cases.
This implies $\bra{\BB^2}=2B_0^2=0.2$ and $1.28$, and therefore $\Brms=0.45$ and $1.13$, respectively.
The initial values of the Alfv\'en speed, $\vAz=\Brms/\sqrt{\mu_0\rho_0}$, are therefore transonic.
We often give the time in code units, $(\cs k_1)^{-1}$, but sometimes we also give it in units of $(\vAz k_0)^{-1}$,
which is physically more meaningful.
However, we must remember that the actual magnetic field and therefore the actual Alfv\'en speed are of course decaying.

In addition to the Roberts field, we add to the initial condition Gaussian-distributed noise of a relative amplitude of $10^{-6}$.
This allows us to study the stability of the field to small perturbations.
To measure the growth rate, we compute the semilogarithmic derivative of $\bra{\JJ_{\perp\perp}^2}/\bra{\JJ^2}$ for a suitable time interval.

The number of eddies in the plane is characterized by the ratio $k/k_1$.
The aspect ratio of the domain is quantified by $L_\|/L_\perp$.
The electric conductivity is quantified by the Lundquist number $\Lu=\vAz/\eta k_0$,
and the kinematic viscosity is related to $\eta$ through the magnetic Prandtl number, $\Pm=\nu/\eta$.
In all our cases, we keep $\Pm=5$.
For most of our runs, we use $\eta\times10^{-7}\cs/k_1$ in code units.

Important output parameters are the growth rate $\lambda=\dd\ln(\bra{\JJ_{\perp\perp}^2}/\bra{\JJ^2})/\dd t$, evaluated in the regime where it is approximately constant.
It is made nondimensional through the combination $\lambda/\vAz k_0$.
We also present magnetic energy and magnetic helicity variance spectra, $\Sp(\BB)$ and $\Sp(h)$, respectively.
These spectra also depend on $k$ and $t$, so we denote the spectra sometimes also as $\Sp(\BB;k,t)$ and $\Sp(h;k,t)$, respectively.

Since $\rho\approx\rho_0=1$, the value of $\Brms$ is also equal to the instantaneous Alfv\'en speed, $\vA$, and its square is the mean magnetic energy density, $\EEM=\bra{\BB^2}/2$.
The latter can also be computed from the magnetic energy spectrum $\EM(k,t)=\Sp(\BB)$ through $\EEM=\int\EM(k,t)\,\dd k$.
The integral scale of the magnetic field is given by
\begin{equation}
\xiM(t)=\int k^{-1}\EM(k,t)\,\dd k/\EEM.
\end{equation}
It is also of interest to compare its evolution with the magnetic Taylor microscale, $\xiT=\Brms/\Jrms$, where $\Jrms$ is the root-mean-squared current density, i.e., $(\nab\times\BB)_\mathrm{rms}$.
(We recall that the permeability was set to unity; otherwise, there would have been an extra $\mu_0$ factor in front of $\Jrms$.)
Both in experiments and in simulations, $\xiT$ may be more easily accessible than $\xiM$, so it is important to find out whether the two obey similar scaling relations.

During the decay, $\EEM=\vA^2/2$ decreases and $\xiM$ increases.
The Alfv\'en time, i.e., the ratio $\tauA\equiv\xiM/\vA$, therefore also increases; see \cite{BJ04} and \cite{HS23} for early considerations of this point.
Both for standard (isotropic) helical decay with $\vA\propto t^{-1/3}$ and $\xiM\propto t^{2/3}$, as well as for nonhelical decay with 
$\vA\propto t^{-5/9}$ and $\xiM\propto t^{4/9}$, the value of $\tauA$ increases linearly with $t$, i.e.,
\begin{equation}
t\propto\tauA(t).
\end{equation}
This is also consistent with the idea that the turbulent decay is self-similar \citep{BK17}.
It was found that the ratio $t/\tauA(t)$ approaches a constant that increases with the Lundquist number \citep{Bra+24}.
The difference between the quantity $t/\tauA(t)$ and the factor $\CM$ defined in \cite{Bra+24}
is the exponent $p=10/9$ in the relation $\EEM\propto t^{-p}$ for nonhelical and $p=2/3$ for helical turbulence with $t/\tauA=\CM/p$.

To compute the Hosking integral, we need the function $\mathcal{I}_\mathrm{H}(R,t)$, which is a weighted integral over $\Sp(h)$, given by
\begin{equation}
\mathcal{I}_\mathrm{H}(R,t)=\int_0^\infty w(k,R)\, \Sp(h;k,t) \, \mathrm{d}k,
\quad \mbox{where} \quad
w(k,R)=\frac{4\pi R^3}{3} \left[\frac{6j_1(kR)}{kR}\right]^2,
\end{equation}
and $j_1(x)=(\sin x-x\cos x)/x^2$ is the spherical Bessel function of order one.
As shown by \cite{Zhou+22}, the function $\mathcal{I}_\mathrm{H}(R,t)$ yields the
Hosking integral in the limit of large radii $R$, although
$R$ must still be small compared with the size of the domain.
They referred to this as the box-counting method for a spherical volume
with radius $R$.

\section{Results}
\label{Results}

\subsection{Isotropization}

\begin{figure}\begin{center}
\includegraphics[width=.49\columnwidth]{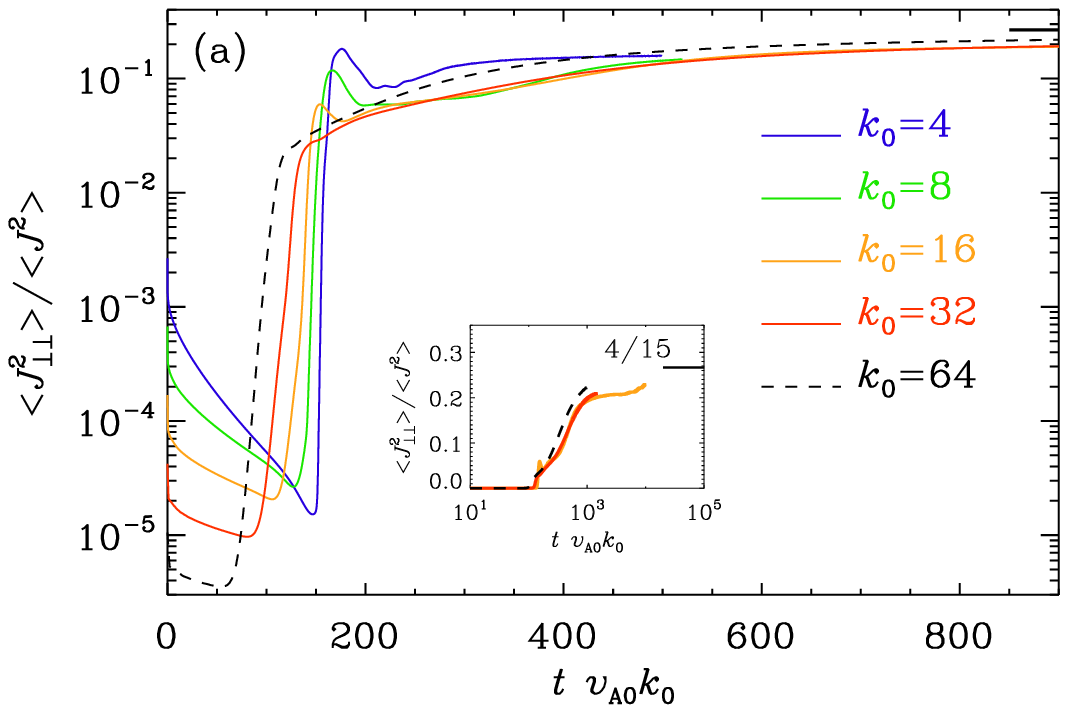}
\includegraphics[width=.49\columnwidth]{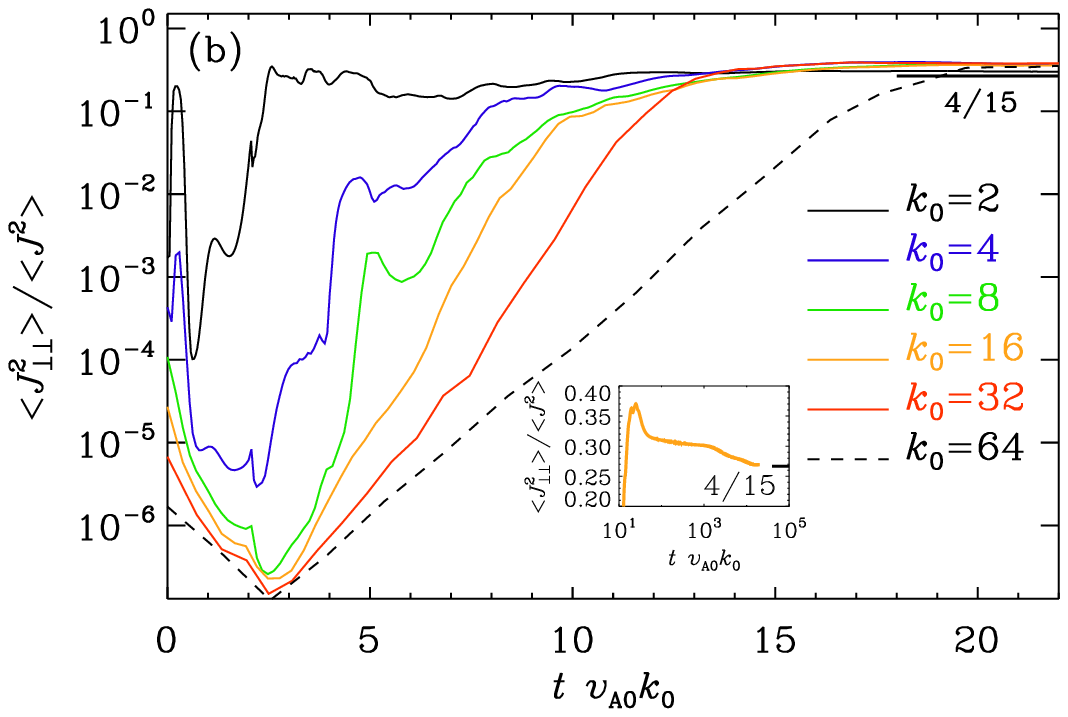}
\end{center}\caption[]{
Evolution of $\bra{\JJ_{\perp\perp}^2}/\bra{\JJ^2}$ for (a) Roberts field I
with $k_0=4$ (blue), $8$ (green), $16$ (orange), $32$ (red), and $64$ (black dashed), and
for (b) Roberts field II with $k_0=2$ (black), $4$ (blue), $8$ (green), $16$ (orange), $32$ (red), and $64$ (black dashed).
The short thick line on the upper right indicates the value of 4/15, which is reached only at much later times outside this plot.
The insets demonstrates that $\bra{\JJ_{\perp\perp}^2}/\bra{\JJ^2}\to4/15$ much later.
}\label{pcomp_plam}\end{figure}

In \Fig{pcomp_plam}, we show the evolution of $\bra{\JJ_{\perp\perp}^2}/\bra{\JJ^2}$ for Roberts fields I and II.
We see that, after a short decay phase, exponential growth commences followed by a saturation of this ratio.
We expect the ratio $\bra{\JJ_{\perp\perp}^2}/\bra{\JJ^2}$ to reach the value $4/15$ at late times; see \App{IsotropicTurbulence}.
The insets of \Fig{pcomp_plam} show the degree to which this is achieved at late times.
Especially in the helical case, when inverse cascading is strong, the peak of the spectrum has already reached the lowest wavenumber of the domain.
This is probably the reason why the value of 4/15 has not been reached by the end of the simulation.

The early growth of $\bra{\JJ_{\perp\perp}^2}/\bra{\JJ^2}$ shows that both the Roberts fields~I and II are unstable to perturbations and develop an approximately isotropic state.
The normalized growth rates are given in \Tab{Tresults} along with the times $\tp$ of maximum growth.
The normalized values are in the range 0.7 to 6, but mostly around unity for intermediate values around $k_0=16$.
The normalized times, $\tp\vAz k_0$, tend to decrease with increasing values of $k_0$ and are about ten to twenty times larger for field~I than for field~II.
This difference was also found in another set of simulations in which $B_0$ was the same for fields~I and II; see \App{DiagnosticDifferentK0}.

\begin{table}\caption{
Normalized growth rates $\lambda$ and peak times $\tp$ for different values of $k_0/k_1$.
The hyphen indicates that no growth occurred.
}\vspace{12pt}\centerline{\begin{tabular}{cccccccc}
field & $k_0=$              &   2  &   4  &  8   &  16  &  32  &  64  \\ 
\hline
 I    & $\lambda/\vAz k_0=$ & ---  & 2.9  & 1.4  & 1.1  & 0.7  & 0.5  \\
II    & $\lambda/\vAz k_0=$ & 5.5  & 1.2  & 0.8  & 1.9  & 1.6  & 1.0  \\
\hline
 I    & $\tp\vAz k_0=$      & ---  &  34  &  16  &  7.7 &  3.4 &  1.2 \\
II    & $\tp\vAz k_0=$      & 1.0  &  1.6 &  2.0 &  0.3 &  0.2 &  0.1 \\
\label{Tresults}\end{tabular}}\end{table}

\begin{figure}\begin{center}
\includegraphics[width=\columnwidth]{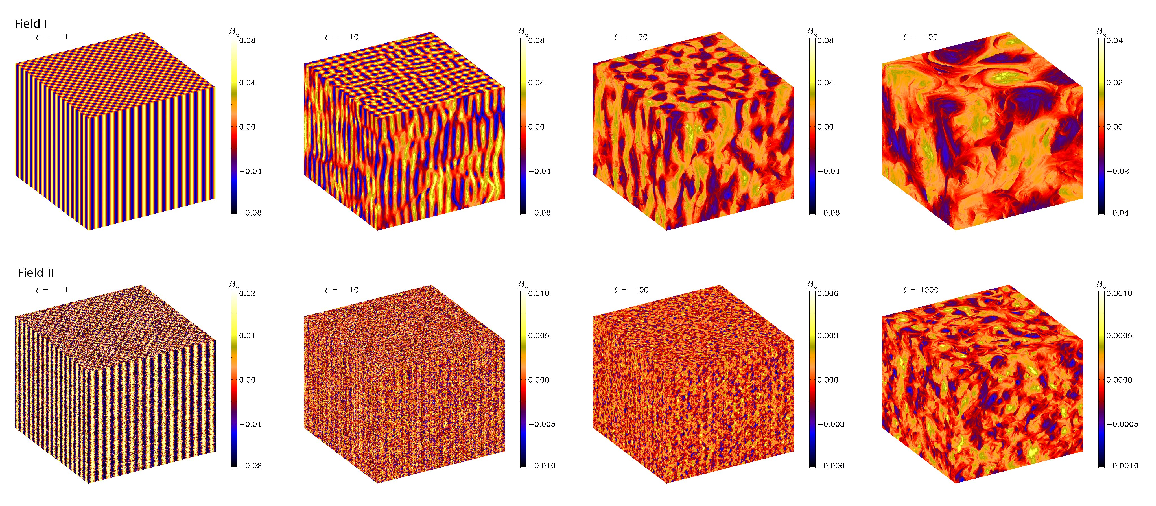}
\end{center}\caption[]{
Visualizations of $B_z$ on the periphery of the computational domain
at times $t=1$, 10, 30, and 100 for Roberts field~I (top) and
at times $t=1$, 10, 100, and 1000 for Roberts field~II (bottom).
}\label{pviz}\end{figure}

Visualizations of $B_z$ on the periphery of the computational domain
are shown in \Fig{pviz} for Roberts fields~I and II.
The initially tube-like structures are seen to decay much faster
for Roberts field~II.
At time $t=100$, the magnetic field has much larger structures for Roberts field~I than at time $t=1000$ for Roberts field~II.

\subsection{Spectral evolution}

\begin{figure}\begin{center}
\includegraphics[width=\columnwidth]{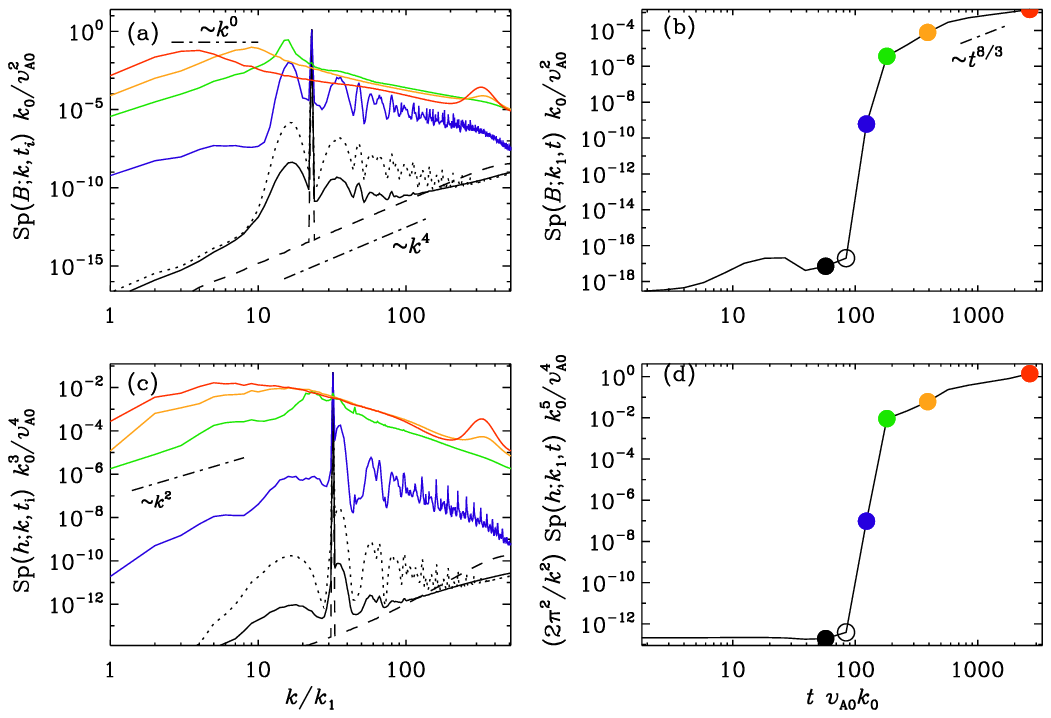}
\end{center}\caption[]{
Evolution of magnetic energy and magnetic helicity variance spectra, $\Sp(\BB)$ and $\Sp(h)$, respectively,
for Roberts field~I with $k_0=16$ at different times $t_i$ indicated by different colors and line types as seen in the time traces on the right.
The open black symbols in panels (b) and (d) correspond to the dotted lines in panels (a) and (c).
}\label{rspec_select_hoskM_1024a2N}\end{figure}

In \Fig{rspec_select_hoskM_1024a2N}, we plot magnetic energy and magnetic helicity variance spectra for the Roberts field~I.
Note that the spectra are normalized by $\vA^2 k_0^{-1}$ and $\vA^4 k_0^{-3}$, respectively.
At early times, the spectra show spikes at $k\approx\kf$ and $2k_0$, respectively, along with higher harmonics.
We also show the time evolution of the normalized values of these spectra at the lowest wavenumber $k=k_1$.
For $\Sp(h)$, we also scale by $2\pi^2/k^2$, which then gives an approximation to the value of the Hosking integral \citep{HS21}.
Again, we see a sharp rise in both time series when the fields becomes unstable.

We also see that at late times, a bump appears in the spectrum near the Nyquist wavenumber.
This shows that the Lundquist number was somewhat too large for the resolution of $1024^3$.
However, comparing with simulations at lower Lundquist numbers shows that the large-scale evolution has not been adversely affected by this.

In \Fig{rspec_select_hoskM_1024b2N}, we show the same spectra for the case of Roberts fields~II.
Again, we see spikes in the spectra at early times.
Those of $\Sp(\BB)$ are again at $\sqrt{2}k_0$, along with overtones, but those of $\Sp(h)$ are now at $2\sqrt{2}k_0$ instead of $2\,k_0$, and there are no spikes of $\Sp(h)$ at $t=0$.
This is a consequence of the fact that the field has zero initial helicity pointwise, and helicity is quickly being produced owing to the growth of the initial perturbations.
The plot of $\Sp(h;k_1,t)$ shows nearly perfectly a constant level for $t\vA k_0=100$.
This indicates that the Hosking integral is well conserved by that time.

\subsection{Spontaneous production of magnetic helicity variance}

\begin{figure}\begin{center}
\includegraphics[width=\columnwidth]{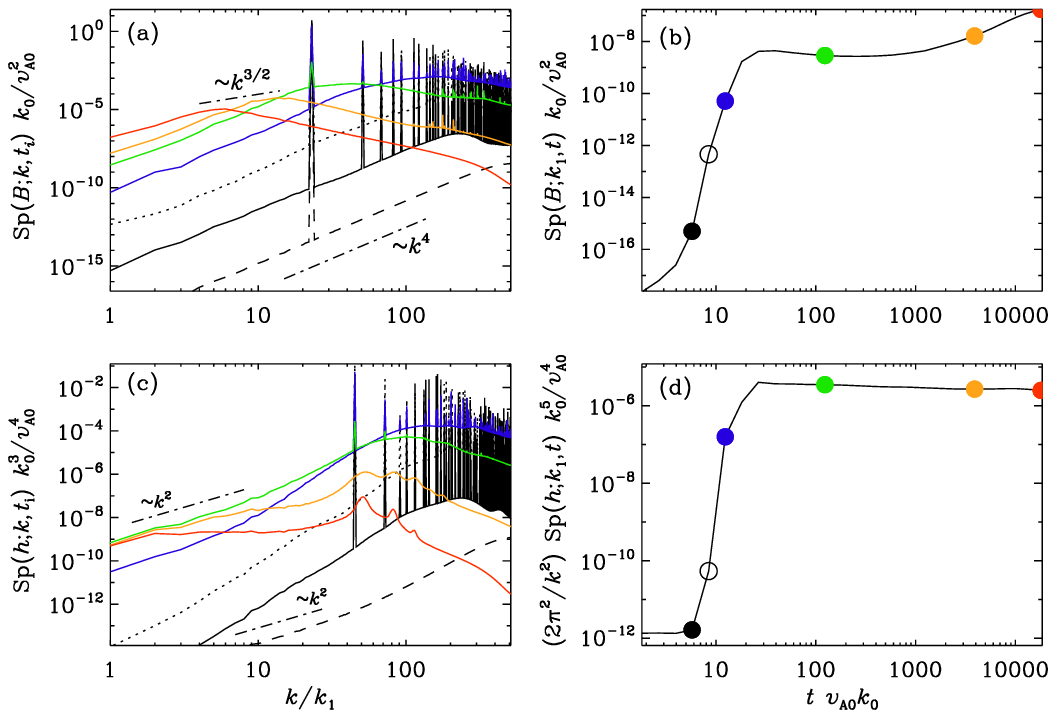}
\end{center}\caption[]{
Same as \Fig{rspec_select_hoskM_1024a2N}, but for the Roberts field~II
at different times $t_i$ as seen in the time traces on the right.
}\label{rspec_select_hoskM_1024b2N}\end{figure}

As we have seen from \Fig{rspec_select_hoskM_1024b2N}, the case of zero magnetic helicity variance is unstable and there is a rapid growth of $\Sp(h)$ also at small wavenumbers.
This was already anticipated by \cite{HS21}, and the present experiments with the Roberts field~II show this explicitly.

\begin{figure}\begin{center}
\includegraphics[width=\columnwidth]{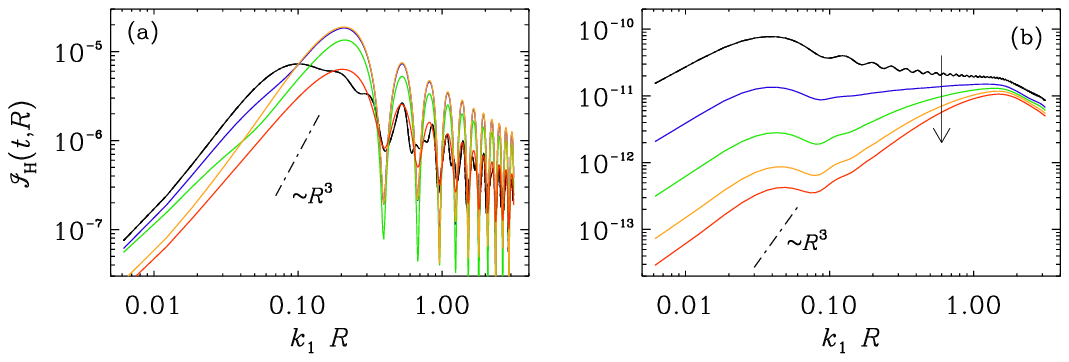}
\end{center}\caption[]{
$\mathcal{I}_\mathrm{H}(R)$ for Roberts field~II with (a) $k_0=4$ 
at $t=1$ (black), 1.5 (blue), 2.2 (green), 3.2 (orange), and 4.6 (red).
and (b) $k_0=16$ at $t=46$ (black), 147 (blue), 316 (green), 570 (orange), and 824 (red).
The arrow indicates the sense of time.
}\label{psaff_comp}\end{figure}

\begin{figure}\begin{center}
\includegraphics[width=\columnwidth]{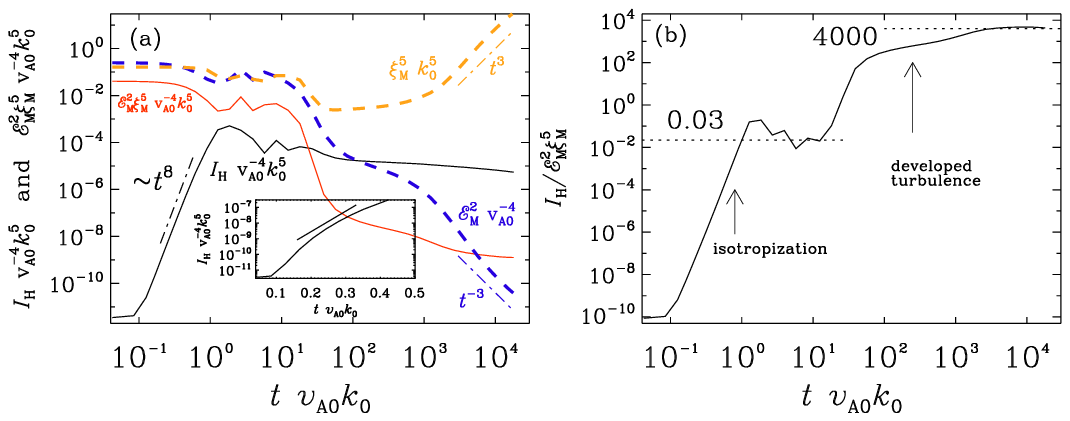}
\end{center}\caption[]{
Time dependence of (a) $I_\mathrm{H}(t)$ (black solid line) along with $\EEM^2\xiM^5$ (red solid line) in units of $\vA^4k_0^{-5}$
as well as $\EEM^2/\vAz^4$ (blue dashed line) and $\xiM^5 k_0^5$ (orange dashed line) and
(b) the ratio $I_\mathrm{H}/\EEM^2\xiM^5$ for Roberts field~II with $k_0=16$.
The plateaus at 0.03 and 3000 are marked by dotted lines.
In (a), the dashed-dotted straight lines indicate the slopes $\propto t^8$ (black),
$\propto t^{3}$ (orange), and $\propto t^{-3}$ (blue).
The inset in (a) shows the growth of $I_\mathrm{H}(t)$ in a semilogarithmic representation along with a line $\propto e^{30 t}$.
}\label{psaff}\end{figure}

We now discuss the function $\mathcal{I}_\mathrm{H}(R,t)$; see \cite{HS21} and \cite{Zhou+22}.
The result is shown in \Fig{psaff_comp}.
For small values of $R$, $\mathcal{I}_\mathrm{H}(R)$ increases $\propto R^3$.
This indicates that the mean squared magnetic helicity density is not randomly distributed on those scales.
In the present case, the actual scaling is slightly shallower than $R^3$, which is probably due to the finite scale separation.
For $R\approx1$, corresponding to scales compatible to the size of the computational domain, we see that $\mathcal{I}_\mathrm{H}(R)$ has a plateau.
It is at those scales, $R=R_\ast$, that we must determine the Hosking integral $I_\mathrm{H}(t)=\mathcal{I}_\mathrm{H}(t,R_\ast)$.
In \Fig{psaff}, we show the time dependence of $I_\mathrm{H}(t)$ for Roberts field II with $k_0=16$ normalized both by $\vAz^4/k_0^5$ (which is constant) and by $\EEM^2\xiM^5$ (which is time-dependent).
Note that the time axis is here also logarithmic.
We see an early rapid growth of $I_\mathrm{H}(t)$ over eight orders of magnitude.

Previous work showed that the value of $I_\mathrm{H}(t)$ can greatly exceed the dimensional estimate $\EEM^2\xiM^5$ \citep{Zhou+22}.
\FFig{psaff} shows that at late times, $t\vAz k_0>100$, this is also the case here.
After the initial rapid growth phase, however, the normalized value of $I_\mathrm{H}(t)$ is still well below unity (around 0.03).
The growth of $I_\mathrm{H}/\EEM^2\xiM^5$ after $t\vAz k_0>100$ is mostly due to the decay of $\EEM$ and it is later counteracted by a growth of $\xiM$.
The dashed blue and orange lines in \Fig{psaff}(a) show separately the evolutions for $\EEM^2/\vAz^4$ and $\xiM^5k_0^5$, respectively.

If the Hosking scaling applies to the present case of initially anisotropic MHD turbulence, we expect $\xiM\propto t^{4/9}$ and therefore $\xiM^5\propto t^{20/9}$.
The actual slope seen in \Fig{psaff} is however around 3 at late times.
For $\EEM$, we expect a $t^{-10/9}$ scaling and therefore $\EEM^2\propto t^{20/9}$, i.e, the reciprocal one of $\xiM^5$.
Again, the numerical data suggest a larger value of around 3.
In \Sec{DiagnosticDiagram}, we analyze in more detail the anticipated scaling of $\EEM(t)\propto t^{-p}$ and $\xiM\propto t^q$.
We find that the two instantaneous scaling exponents $p$ and $q$ are indeed larger than what is expected based on the Hosking phenomenology.
However, the instantaneous scaling exponents also show a clear evolution toward the expected values.

It is interesting to observe that the evolution of $I_\mathrm{H}$ proceeds in two distinct phases.
In the first one, when $t\vAz k_0<2$, $I_\mathrm{H}$ shows a rapid growth that is not exponential; see the inset of \Fig{psaff},
where the growth of $I_\mathrm{H}$ is shown on a semilogarithmic representation.
The growth is closer to that of a power law, at the approximate exponent would be around six, which is rather large.
During this phase, the turbulent cascade has not yet developed, but a nonvanishing and nearly constant value of $I_\mathrm{H}$ has been established.
However, in units of $\EEM^2\xiM^5$, its value is rather small (around 0.03).

In the second phase, when $t\vAz k_0>100$, turbulence has developed, and a turbulent decay has developed.
It is during this time that the ratio $I_\mathrm{H}(t)/\EEM^2\xiM^5$ approaches larger values (around 3000) that were previously seen in isotropic decaying turbulence simulations \citep{Zhou+22}.
The reason for this large value was argued to be due to the development of non-Gaussian statistics in the magnetic field.
However, \cite{BB25} presented an estimate in which the value of this ratio is equal to $\CM^2$.
With $\CM\approx50$, this would agree with the numerical findings.

\section{Cosmological applications}
\label{CosmologicalApplications}

\subsection{Evolution in the diagnostic diagram}
\label{DiagnosticDiagram}

\begin{figure}\begin{center}
\includegraphics[width=\columnwidth]{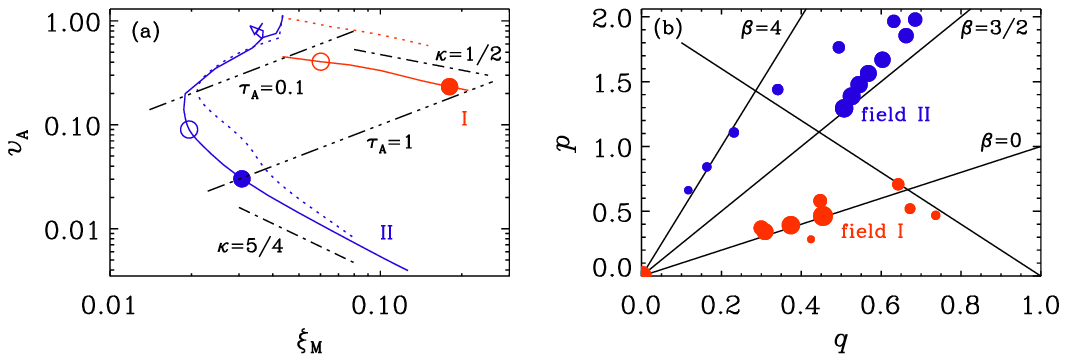}
\end{center}\caption[]{
(a) Parametric representation of $\vA$ versus $\xiM$ for Roberts fields I (red) and II (blue).
The solid (dotted) curves are for $\eta=2\times10^{-7}$ ($\eta=4\times10^{-6}$).
Note that the red dotted line for $\eta=4\times10^{-6}$ starts at the same value $\vA=\sqrt{1.28}$ as the nonhelical runs (blue lines).
The similarity between the dotted and solid red lines shows that the initial amplitude does not matter much.
The open (filled) symbols indicate the times $t=10$ ($t=100$).
The dashed-dotted lines give the slopes $\kappa=1/2$ and 5/4 for Roberts fields~I (red) and II (blue), respectively.
(b) $pq$ diagram field fields~I (red) and II (blue) with $\eta=2\times10^{-7}$.
Larger symbols indicate later times.
}\label{pq}\end{figure}

In view of the cosmological applications of decaying MHD turbulence, it is of interest to consider the evolution of the actual Alfv\'en speed $\vA(t)=\sqrt{2\EEM/\rho}$
in an evolutionary diagram as a parametric representation versus $\xiM(t)$; see \Fig{pq}(a).
With $\vA\propto t^{-p/2}$ and $\xiM\propto t^{q}$, we expect that $\vA\propto\xiM^{-\kappa}$ with $\kappa=p/2q=1/2$ for the fully helical case of Roberts field~I.
This is in agreement with early work showing that $\vA\propto t^{1/3}$ and $\xiM\propto t^{2/3}$ \citep{Hat84,BM99}.

\begin{figure}\begin{center}
\includegraphics[width=\columnwidth]{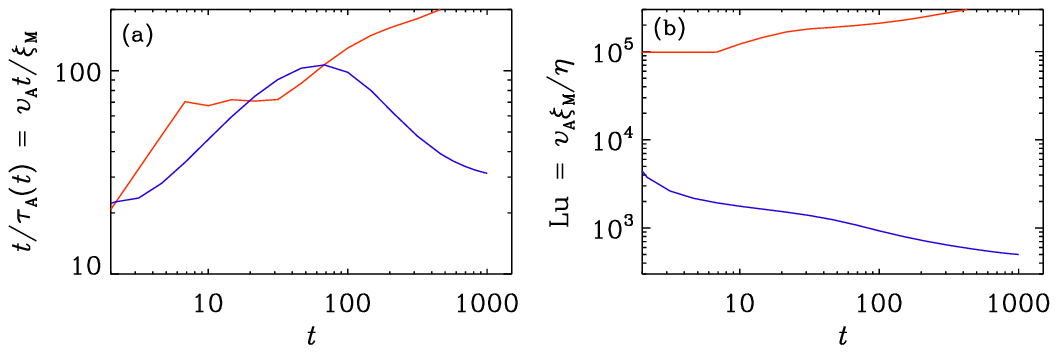}
\end{center}\caption[]{
(a) $t/\tauA$ and (b) $\Lu$ versus time for Roberts fields I (red) and II (blue).
}\label{pcomp_jbrms_II_cm}\end{figure}

In \Fig{pq}(a), we have also marked the times $t=10$ (open symbols) and $t=100$ (filled symbols).
The points of constant times depart significantly from the lines of constant Alfv\'en time, $\tau_\mathrm{A}$, for which $\vA=\xiM/\tau_\mathrm{A}$ grows linearly with $\xiM$.
We expect the times to be larger by a factor $\CM$ than the corresponding values of $\tau_\mathrm{A}(t)$.
This is indeed to case: the point $t=100$ lies on the line $\tauA=1$, i.e., $t/\tauA=100$.
This is twice as much as our nominal value of about 50.

There is an interesting difference between the cases of Roberts fields~I and II in that for field~II,
there is an extended period during which $\xiM$ shows a rapid decrease before the expected increase emerges.
The fact that such an initial decrease of the characteristic length scale is not seen for Roberts field~I is remarkable.
The rapid development of smaller length scales is probably related to the breakup of the initially organized tube-like structures into smaller scales.
In the helical case, however, the nonlinear interaction among helical modes can only result in the production of modes with smaller wavenumbers, i.e., larger length scales;
see \cite{Frisch+75} and \cite{BS05} for a review.
Such a constraint does not exist for the nonhelical modes, where
this can then reduce the \textit{effective} starting values of $\xiM$ and therefore also of the effective Alfv\'en time, $\tauA=\xiM/\vA$, early in the evolution.
In \App{DiagnosticDifferentK0}, we present similar diagrams for different values of $k_0$, but with a drag term included that could be motivated by cosmological applications.

We inspect the time-dependences of $t/\tauA=\vA t/\xiM$ and $\Lu=\vA\xiM/\eta$ for Roberts fields~I and II in \Fig{pcomp_jbrms_II_cm}.
We see that $t/\tau_\mathrm{A}(t)$ reaches values in excess of 100 for $t=100$ in both cases.
This is more than what has been seen before, but it also shows significant temporal variations.

\begin{figure}\begin{center}
\includegraphics[width=.49\columnwidth]{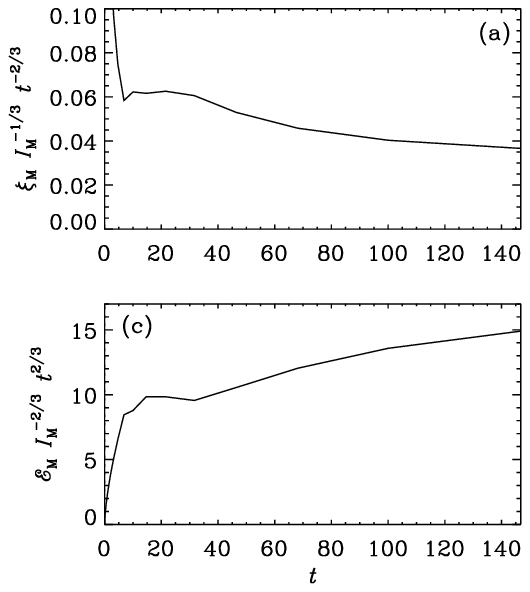}
\includegraphics[width=.49\columnwidth]{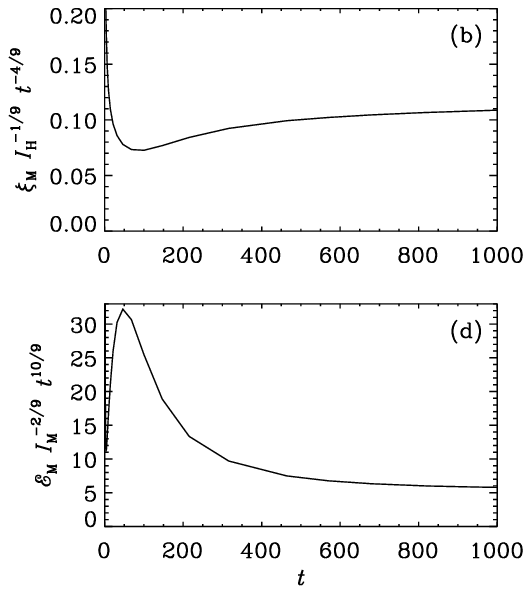}
\end{center}\caption[]{
Compensated evolutions of $\xiM$ and $\EEM$ allowing
the nondimensional prefactors in \Eq{Universal} to be estimated.
}\label{phel}\end{figure}

\begin{table}\caption{
Summary
}\vspace{12pt}\centerline{\begin{tabular}{lcccccc}
Reference & $C_\mathrm{M}^{(\xi)}$ & $C_\mathrm{H}^{(\xi)}$ & $C_\mathrm{M}^{({\cal E})}$ & $C_\mathrm{H}^{({\cal E})}$ & $C_\mathrm{M}^{(E)}$ & $C_\mathrm{H}^{(E)}$ \\
\hline
\cite{BB25}  & 0.12 & 0.14 & 4.3 & 4.0 & 0.7 & 0.025 \\
\cite{BSV23} & ---  & 0.12 & --- & 3.7 & --- & 0.025 \\
\cite{BL23}  & ---  & 0.15 & --- & 3.8 & --- & 0.025 \\
present work & 0.04 & 0.10 & 15  &  6  & --- &  ---  \\
\label{TOverview}\end{tabular}}\end{table}

\subsection{Universality of prefactors in decay laws?}

The decay of a turbulent magnetic field is constrained by a certain conservation laws:
the conservation of mean magnetic density $I_\mathrm{M}=\bra{h}$, where $h=\AAA\cdot\BB$ is the local magnetic helicity density,
and the Hosking integral, $I_\mathrm{H}=\int h(\xx)h(\xx+\rr)\,\dd^3r$.
When the magnetic field is fully helical, the decay is governed by the conservation of $I_\mathrm{M}$, and when it is nonhelical, it is governed by the conservation of $I_\mathrm{H}$.
The time of crossover depends on the ratio $t_\ast\equiv I_\mathrm{H}^{1/2}/I_\mathrm{M}^{3/2}$ \citep{BB25}.
Specifically, the correlation length $\xiM(t)$, the mean magnetic energy $\EEM(t)$, and the envelope of the peaks of the magnetic energy spectrum $\EM(k,t)$ depends on the values
of the conserved quantities with \citep{BL23}
\begin{equation}
\xiM(t)=C_i^{(\xi)} I_i^\sigma t^q,\quad
\EEM(t)=C_i^{({\cal E})} I_i^{2\sigma} t^{-p},\quad
\EM(k)\leq C_i^{(E)} I_i^{(3+\beta)\,q} k^\beta.
\label{Universal}
\end{equation}
where $\sigma$ is the exponent with which the length enters in $I_{i}$: $\sigma=3$ when the mean magnetic helicity density governs the decay ($i={\rm M}$)
and $\sigma=9$ for the Hosking integral ($i={\rm H}$).
In \Fig{phel}, we show the appropriately compensated evolutions of $\xiM$ and $\EEM$ such that we can read off the values of $C_i^{(\xi)}$ and $C_i^{({\cal E})}$ for the helical and nonhelical cases.

In \Tab{TOverview}, we summarize the values for the six coefficients reported previously in the literature and compare with those determine here.
The fact that the coefficients are now somewhat different under different circumstances suggests that they might not be universal,
although the more complicated setup of the present experiment as well as limited scale separation may have contributed to their present results.
This question is significant, however, because universality would mean that the decay laws of the form \citep[e.g.,][]{Vac21}
\begin{equation}
\xiM(t)=\xiM(t_0)\,\left(t/t_0\right)^q,\quad
\EEM(t)=\EEM(t_0)\,\left(t/t_0\right)^{-p}
\end{equation}
could be misleading in that they suggest some freedom in the choice of the values of $\xiM(t_0)$ and $\EEM(t_0)$ at the time $t_0$.
Comparing with \Eq{Universal}, we see that
\begin{equation}
\xiM(t_0)/t_0^q=C_i^{(\xi)} I_i^\sigma,\quad\mbox{and}\quad
\EEM(t_0)\,t_0^p=C_i^{({\cal E})} I_i^{2\sigma},
\end{equation}
so they cannot be chosen arbitrarily, but they must obey a constraint that depends on the relevant conservation law.

\section{Conclusions}
\label{Conclusions}

We have seen that a tube-like arrangement of an initial magnetic field becomes unstable to small perturbations.
The resulting magnetic field becomes turbulent and tends to isotropize over time.
This means that tube-like initial conditions that could be expected in plasma experiments would allow us to study the turbulent MHD decay dynamics -- even for moderate but finite scale separation
of 4:1 or more.
In other words, the number of tubes per side length should be at least four.

We have also seen that a pointwise nonhelical magnetic field, as in the case of the Roberts field II, is unstable and develops magnetic helicity fluctuations.
After about one Alfv\'en time, the Hosking integral reaches a finite value, but a fully turbulent decay commences only after about one hundred Alfv\'en times.
From that time onward, the value of the Hosking integral relative to that expected on dimensional grounds reaches a value of several thousand, a value that was also found earlier \citep{Zhou+22}.

Our present results have confirmed the existence of a resistively prolonged turbulent decay time whose value exceeds the Alfv\'en time by a factor $\CM\approx\tau/\tauA$.
As emphasized above, the fact that this ratio depends on the microphysical magnetic diffusivity is in principle surprising, because one of the hallmarks of turbulence
is that its macroscopic properties should not depend on the microphysics of the turbulence.
It would mean that it is not possible to predict this behavior of MHD turbulence
by ignoring the microphysical magnetic diffusivity, as is usually done in so-called large eddy simulations.

The present results have shown that the decay time can exceed the Alfv\'en time by a factor of about, which is similar to what was found previously \citep{Bra+24}.
During intermediate times, however, the decay time can even be a hundred times longer than the Alfv\'en time.
The dimensionless prefactors in the dimensionally motivated powerlaw expressions for length scale and mean magnetic energy density are also roughly similar to what was previously
obtained from fully isotropic turbulence simulations.


\section*{Funding}
This work was supported in part by the Swedish Research Council (Vetenskapsr{\aa}det, 2019-04234),
the National Science Foundation under grant no.\ NSF AST-2307698 and a NASA ATP Award 80NSSC22K0825.
National Key R\&D Program of China (No.\ 2021YFA1601700), and the National Natural Science Foundation of China (No.\ 12475246).
We acknowledge the allocation of computing resources provided by the
Swedish National Allocations Committee at the Center for Parallel
Computers at the Royal Institute of Technology in Stockholm and
Link\"oping.

\section*{Declaration of Interests}
The authors report no conflict of interest.

\section*{Data availability statement}
The data that support the findings of this study are openly available on
\url{http://norlx65.nordita.org/~brandenb/projects/Roberts-Decay/}.
All calculations have been performed with the {\sc Pencil Code}
\citep{JOSS}; DOI:10.5281/zenodo.3961647.

\section*{Authors' ORCIDs}

\noindent
A. Brandenburg, https://orcid.org/0000-0002-7304-021X

\noindent
Longqing Yi, https://orcid.org/0000-0003-2378-5480

\noindent
Xianshu Wu, https://orcid.org/0000-0003-3382-5132

\appendix

\begin{figure}\begin{center}
\includegraphics[width=\columnwidth]{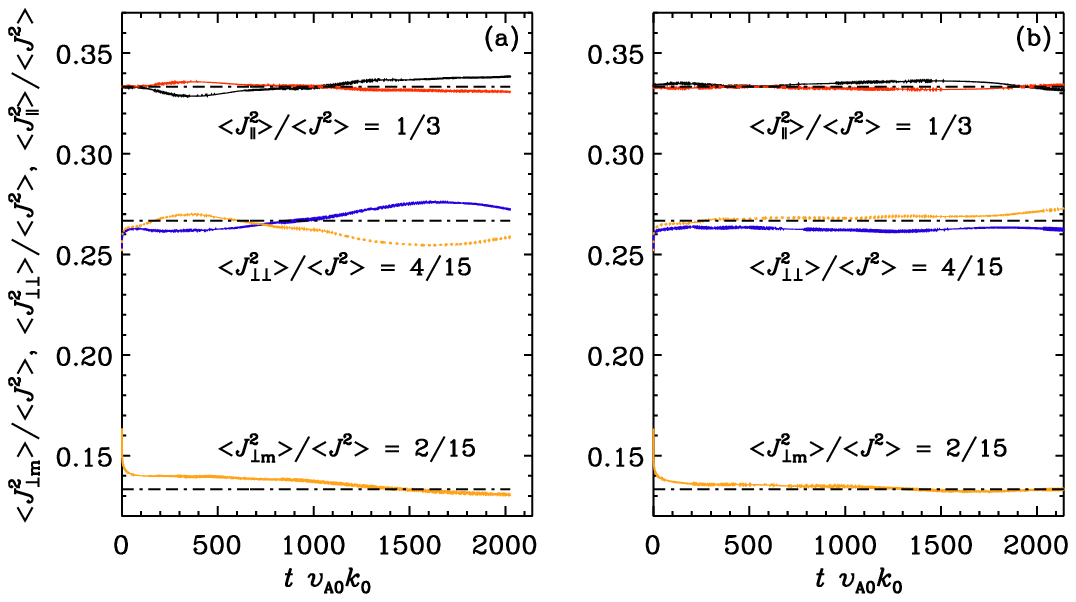}
\end{center}\caption[]{
Evolution of $\bra{\JJ_{\perp\mathrm{m}}^2}/\bra{\JJ^2}$,
$\bra{\JJ_{\perp\perp}^2}/\bra{\JJ^2}$, and $\bra{\JJ_{\|}^2}/\bra{\JJ^2}$
for decaying isotropic turbulence with an initial peak wavenumber $k_0/k_1=8$
using $1024^3$ meshpoints (a) with helicity and (b) without helicity.
}\label{piso}\end{figure}

\section{$\bra{\JJ_{\perp\perp}^2}/\bra{\JJ^2}$ for isotropic turbulence}
\label{IsotropicTurbulence}

We have examined the evolution of $\bra{\JJ_{\perp\perp}^2}/\bra{\JJ^2}$ for isotropic turbulence using a setup similar to that of \cite{BSV23}; see \Fig{piso}.
The scale separation, i.e., the ratio of the peak wavenumber to the lowest wavenumber in the domain is 8 for this simulation and the Lundquist number,
which is the rms Alfv\'en speed times the correlation length divided by the magnetic diffusivity, is about $10^4$.
The other parameters are as in the earlier work of \cite{BSV23}; see the data availability statement of the present paper.

\section{Diagnostic diagrams for different $k_0$}
\label{DiagnosticDifferentK0}

\begin{figure}\begin{center}
\includegraphics[width=.49\columnwidth]{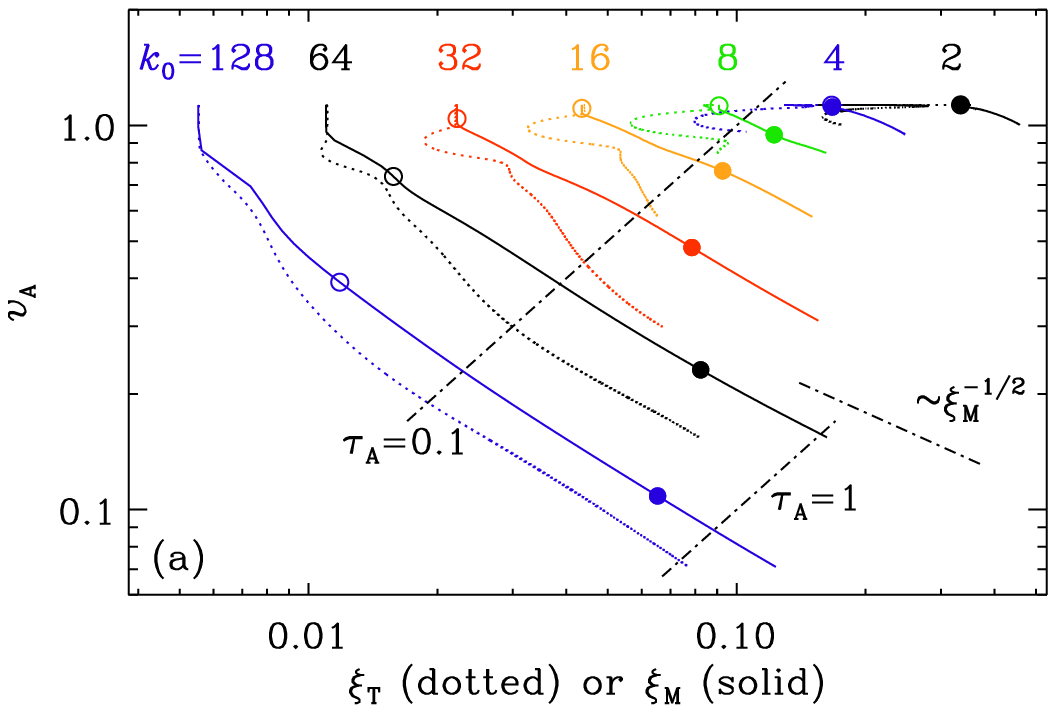}
\includegraphics[width=.49\columnwidth]{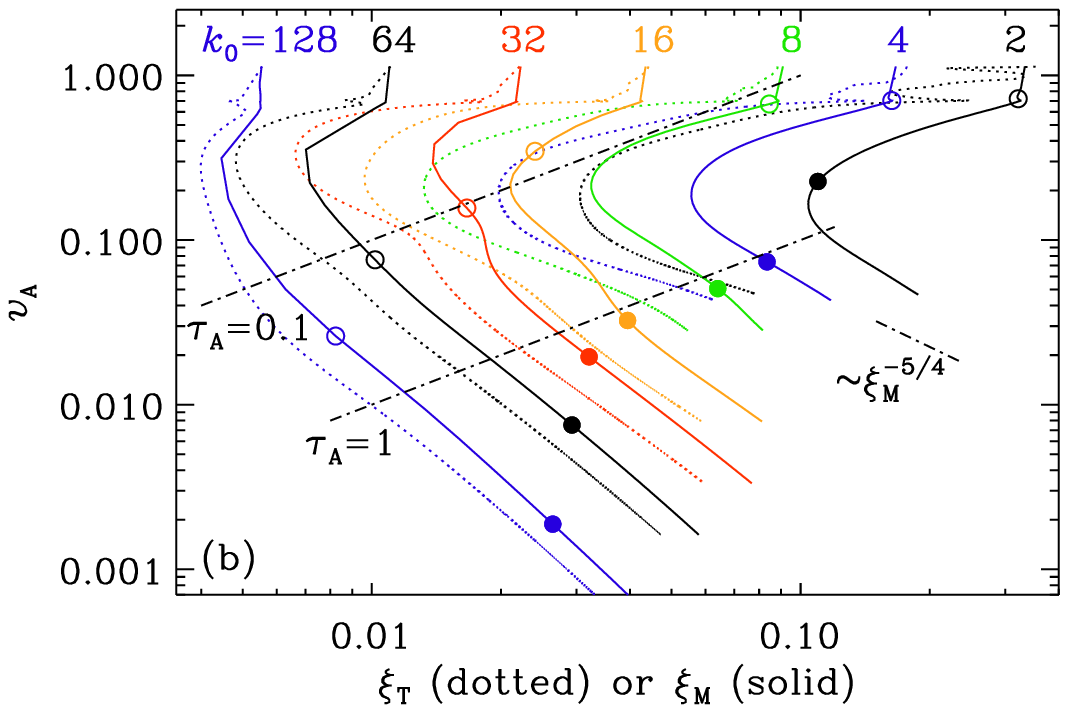}
\end{center}\caption[]{
Same as \Fig{pq}(a), but for $c_\alpha=3$, showing a
parametric representation of $\Brms$ versus $\Brms/\Jrms$ and $\xiM$ for Roberts field I (left)
with $k_0=2$ (black), $4$ (blue), $8$ (green), and $16$ (orange),
$32$ (red), $64$ (black), and 128 (blue).
The open (filled) symbols in both plots indicate the times $t=10$ ($t=100$).
}\label{pcomp_jbrms_alp}\end{figure}

In \Fig{pq}, we did already present a diagnostic diagrams of $\vA$ vs.\ $\xiM$ for $k_\mathrm{p}=16$.
We also performed runs for different values of $k_\mathrm{p}$ to compute the growth rates and the times $\tp$ of maximum growth in \Tab{Tresults},
but not all the runs were long enough to compute similar tracks in the diagnostic diagram.
In \Fig{pcomp_jbrms_alp}, we show such a diagram for a case in which a drag term of the form $-\alpha\uu$ is included on the right-hand side of \Eq{DuDt}.
Here, we choose a drag coefficient that automatically changes in time so as to allow for a nearly self-similar decay.
Using a multiple of $1/t$ is an obvious possibility, but it would always be the same at all locations and for different types of flows.
The local vorticity might be one possible option for a coefficient that varies in space and time, and has the right dimension.
Another possibility, which is also the one chosen here, is to take $\alpha$ to be a multiple of $\sqrt{\mu_0/\rho_0}|\JJ|$ and write $\alpha=c_\alpha \sqrt{\mu_0/\rho_0}|\JJ|$,
where $c_\alpha$ is a dimensionless prefactor, and $\mu_0=\rho_0=1$ has been set.
Again, as was already clear from \Fig{pq}, the tracks without helicity show a marked excursion to smaller values of $\xiM$ before displaying a decay of the form $\vA\propto\xiM^{-\kappa}$.
The corresponding values of $\lambda/\vAz k_0$ and $\tp\vAz k_0$ are given in \Tab{Tresults_alp}.

\begin{table}\caption{
Similar to \Tab{Tresults}, showing normalized growth rates $\lambda$ and peak times $\tp$ for different values of $k_0$, but with the photon drag term included.
Here, unlike the case of \Tab{Tresults}, the values of $B_0$ are the same for Roberts fields~I and II.
The hyphen indicates that no growth occurred.
The lowest value of $k_0$ has been set in italics to indicate that we used here what we called the rotated Roberts field.
}\vspace{12pt}\centerline{\begin{tabular}{crcccccccc}
field & $k_0=$ & \textit{0.71} &   1  &   2  &   4  &  8   &  16  &  32  &  64  \\ 
\hline
 I    & $\lambda/\vAz k_0=$ & ---  & ---  & 0.01 & 0.02 & 0.05 & 0.05 & 0.05 & 0.05 \\
II    & $\lambda/\vAz k_0=$ & 0.12 & 0.15 & 0.19 & 0.20 & 0.22 & 0.22 & 0.19 & 0.13 \\
\hline
 I    & $\tp\vAz k_0=$      & ---  & ---  & 310  & 122  & 62   &  31  &  12  &  4.5 \\
II    & $\tp\vAz k_0=$      &  78  &  51  &  27  &  14  &  6.7 &  3.5 &  1.8 &  1.2 \\
\label{Tresults_alp}\end{tabular}}\end{table}

Our definition of the Roberts fields follows the earlier work by \cite{Rhei+14}.
In the original paper by \cite{Rob72}, however, the field was rotated by $45\degr$.
In that case, $\phi=\cos k_0 x \mp\cos k_0 y$, where the upper and lower signs refer to Roberts fields~I and II.
For this field, a lower eigenvalue of the curl operator, namely $\kf=k_0$, can be accessed.
In that case, we can accommodated 1 pair of flux tubes instead of four.
This can be done both for fields~I and II.
They are given by
\begin{equation}
\BB_\mathrm{I}=\pmatrix{
\sin k_0 y\cr
\sin k_0 x\cr
\cos k_0 x -\cos k_0 y},
\quad
\BB_\mathrm{II}=\pmatrix{
\sin k_0 y\cr
\sin k_0 x\cr
\cos k_0 x +\cos k_0 y},
\end{equation}
which satisfies $\BB_\mathrm{I}\cdot\nab\times\BB_\mathrm{I}=\kf\BB_\mathrm{I}^2$
and $\BB_\mathrm{II}\cdot\nab\times\BB_\mathrm{II}=0$, just like the nonrotated field.
But here, $\kf=k_0$ is the eigenvalue of the curl operator.

\bibliography{ref}{}
\bibliographystyle{jpp}
\end{document}